%% file: main.tex
\documentclass[10pt, conference, compsocconf]{IEEEtran}

\usepackage{mathptm}
\usepackage[utf8]{inputenc}
\usepackage[T1]{fontenc}
\usepackage{babel}
\usepackage{textcomp}
\usepackage{type1cm}
\usepackage{amsmath}
\usepackage{amsfonts}
\usepackage{subcaption}
\usepackage{booktabs}
\usepackage{listings}
\usepackage{graphicx}
\let\labelindent\relax
\usepackage{enumitem}
\usepackage{color}

\usepackage{multirow}
\usepackage[symbol]{footmisc}
\usepackage{refcount}
\usepackage{scrextend}
\usepackage{float}

\DeclareMathOperator*{\argext}{arg\,ext}
\DeclareMathOperator*{\argextP}{arg\,ext_P}

\DeclareMathOperator*{\extP}{ext_P}
\begin{document}


\title{Analyzing Search Techniques for Autotuning Image-based GPU Kernels:\\The Impact of Sample Sizes}


\author{
\IEEEauthorblockN{Jacob O. Tørring and Anne C. Elster}
\IEEEauthorblockA{Department of Computer Science\\
Norwegian University of Science and Technology (NTNU)\\
Trondheim, Norway\\
Email: \{jacob.torring, elster\}@ntnu.no
}
}

\maketitle

\input{content/abstract}


%
\IEEEpeerreviewmaketitle

\input{content/introduction/introduction_2.tex}
\input{content/background/background.tex}
\input{content/related-work/related-work}

\input{content/method/method}
\input{content/results/results.tex}
\input{content/conclusion.tex}
\input{content/acknowledgements}

\input{content/generated/references.tex}	

\end{document}

%% file: content/abstract.tex
\begin{abstract}
Modern computing systems are increasingly more complex, with their multicore CPUs and GPUs accelerators changing yearly, if not more often. It thus has become very challenging to write programs that efficiently use the associated complex memory systems and take advantage of the available parallelism. Autotuning addresses this by optimizing parameterized code to the targeted hardware by \textit{searching} for the optimal set of parameters. Empirical autotuning has therefore gained interest during the past decades. While new autotuning algorithms are regularly presented and published, we will show why comparing these autotuning algorithms is a deceptively difficult task. 

In this paper, we describe our empirical study of state-of-the-art search techniques for autotuning by comparing them on a range of sample sizes, benchmarks and architectures. We optimize 6 tunable parameters with a search-space size of over 2 million. 
The algorithms studied include Random Search (RS), Random Forest Regression (RF), Genetic Algorithms (GA), Bayesian Optimization with Gaussian Processes (BO GP) and Bayesian Optimization with Tree-Parzen Estimators (BO TPE). 

Our results on the ImageCL benchmark suite suggest that the ideal autotuning algorithm heavily depends on the sample size.
In our study, BO GP and BO TPE outperform the other algorithms
in most scenarios with sample sizes from 25 to 100. However, GA usually outperforms the others for sample sizes 200 and beyond. 
We generally see the most speedup to be gained over RS in the lower range of sample sizes (25-100). However, the algorithms more consistently outperform RS for higher sample sizes (200-400).  Hence, no single state-of-the-art algorithm outperforms the rest for all sample sizes. Some suggestions for future work are also included.

\end{abstract}

%% file: content/introduction/introduction_2.tex
\section{Introduction}\label{chap:introduction}
  
 
While CPUs are still the core of modern computers, we have seen a significant increase in their complexity in recent decades. 
Accelerators such as GPUs (Graphics Processing Units) 
can greatly improve efficiency and performance if developers can utilize them correctly. 
Hence, new data
centers, and most of the world's Top500 No. 1 systems since 2010
\cite{elster_european_2021}
%
are typically HPC (High-Performance Computing) systems with large collections of both CPUs and GPUs. 

Increasingly \emph{heterogeneous} systems make the task of optimizing for performance an increasingly difficult one, with rapidly changing GPU / Accelerator architectures complicating matters even further.

\subsection{The case for autotuning}
Researchers spend considerable efforts in computer science and related fields on optimization of their programs, so they can take advantage of the significant yearly improvements in the performance of HPC systems.

%
However, as the complexity of these systems grows, optimizing programs grows beyond the practical limits of developers' ability to understand the entire system. 

To address this, traditionally built analytical methods -- as in the case of compiler optimization -- are used to understand how all the system components relate to one another. The knowledge of these systems is then turned into \emph{rules} that can be followed to optimize programs. The rules can add, modify or remove large chunks of code without changing the \emph{semantics} of the program. However, sometimes these rules cannot encode the optimal solution perfectly, and so they contain  \emph{heuristics}. 

While heuristics often perform well, they can also limit the program from finding the optimal solution, especially in ever-changing heterogeneous architectures. In fact, sometimes new inventive methods that work well for their time can be shown later not to be ideal \cite{meyer_latency_2008}.

To find the optimal solution, one therefore often needs to test the programs and observe how they perform. Using this additional data, one can apply \emph{empirical methods} that \emph{search} for the optimal solution through trial and error. When these 
empirical methods are applied to general programs, we refer to the methods as \emph{autotuning}.


Ideally, compilers and translators should hide as many architecture-based optimization details as possible from the language user so that their codes can be used efficiently across generations of systems. This was one of the key consideration when Falch and Elster developed the ImageCL programming language~\cite{falch_imagecl_2016} that was built upon OpenCL~\cite{stone_opencl_2010}
. To abstract away several of the GPU-specific parameters without any input needed from the end-user, they developed an \emph{autotuner} 
~\cite{falch_machine_2017} that showed promising results.


In this paper, we compare several state-of-the-art autotuning techniques that are commonly used in practice across various benchmarks and architectures. 
These include
Random Forest Regression (RF), a traditional technique that was used in the seminal work by James Bergstra~\cite{bergstra_machine_2012}. 
Sequential model-based optimization techniques like Bayesian Optimization with Gaussian Processes (BO GP)~\cite{snoek_practical_2012} and 
Tree-Parzen Estimators (BO TPE)~\cite{bergstra_algorithms_2011} have also seen widespread adoption~\cite{bergstra_hyperopt_2013, hutter_sequential_nodate}. 
In addition to these techniques, 
\emph{metaheuristic} algorithms 
are also commonly used
such as Particle Swarm Optimization (PSO), Simulated Annealing (SA) and Genetic Algorithms(GA)~\cite{nugteren_cltune:_2015, van_werkhoven_kernel_2019}.
However, despite extensive research into new autotuning methods, there is still no consensus across these disciplines on which autotuning method is optimal. 

Our goals are to investigate whether there is an optimization algorithm that outperforms all others for all sample sizes and how the optimal selection of autotuning algorithms changes as a function of benchmark, hardware, and sample size. This work is an extension of previous work~\cite{torring_optimization_2020}.

The rest of this paper will be organized as follows:
    The following section 
    Section~\ref{chap:background} 
    introduces general theoretical background,
    followed by an introduction to autotuning techniques (Section~\ref{sec:autotuning}).
    Section~\ref{chap:related-work} introduces and discusses related work, whereas 
    Section~\ref{sec:experimental-design} details our experimental design and Section~\ref{sec:framework} details the framework we use to conduct the experiments.
    Finally, Section~\ref{chap:results} contains our experimental results and our discussion of the data, and 
    Section~\ref{chap:conclusion} concludes the paper and gives some ideas regarding future work left to be done.
   

%% file: content/background/background.tex
\section{Background} \label{chap:background}


GPUs were originally designed for fast processing of graphics. However, in recent decades they have also been found to be excellent accelerators for sufficiently parallelizable workloads. GPUs have thus been used as General Purpose GPUs (GPGPU) to accelerate a wide range of scientific computations. 
Many of the largest computing clusters in the world today use GPUs as an essential computational resource. 
The design and optimization of GPU workloads and architectures have therefore become important for many other subjects than video game graphics.

\subsection{Performance Portability}
Due to the wide range of hardware available for computers, a program will often \emph{functionally} behave the same for all compatible platforms. However, the \emph{performance} can vary greatly. A configuration of a problem that runs well on one GPU might perform terribly on another GPU of a different architecture. 
ATLAS~\cite{clint_whaley_automated_2001} is an Automatically Tuned Linear Algebra library, that is an early well-known effort that addresses this by testing out various configurations of matrix block sizes to decide which configuration is optimal for a given system's memory hierarchy. Similar concepts are 
also part of Intel's matrix library MKL and the FFT library FFTW~\cite{frigo_fftw_1998}. 



\subsection{ImageCL}
ImageCL~\cite{falch_imagecl_2016} is an OpenCL-based language for abstracting some of the hardware-specific details of OpenCL to provide performance portability. The language abstracts away parameters such as the work group size and thread coarsening from OpenCL into \emph{tuning parameters} via a configuration file. \emph{Thread coarsening} describes how many data elements each thread has to process. After the configuration has been specified, the program can be compiled and run to test its performance with the assigned configuration. The ImageCL autotuner AUMA~\cite{falch_machine_2017} uses this process to tune the ImageCL kernels automatically. Designing and implementing a general optimal autotuner is still an open research question.


\input{content/background/statistics-and-comparability}

\input{content/related-work/auto-tuning}


%% file: content/background/statistics-and-comparability.tex
\subsection{Significance tests and effect sizes}
When evaluating program characteristics such as performance, the result will often vary significantly depending on many uncontrollable factors, e.g. OS scheduling, caching, clock frequencies, branch predictors, etc. 
We therefore need statistical tools that can handle this uncertainty and still allow us to be confident in the results. 
This section will introduce our chosen significance test and effect size for our study. 

\subsubsection{Mann-Whitney U test}\label{sec:mwu}
If the variables from the distributions we are comparing can be assumed to be \emph{independent}, we can use the non-parametric Mann-Whitney U test~\cite{mann_test_1947}. This test can be used to investigate whether two independent samples are from populations with identical distributions. The null hypothesis is that a randomly chosen sample from one population is equally likely to be less than or greater than a randomly selected sample from the other population. 

\subsubsection{Common Language Effect Size}
While significance tests can be used to determine whether there \emph{is} a significant difference between populations, it does not say anything about the \emph{size} of the difference. 
We can get a low p-value in two ways, either by having a large difference and few samples, or a small difference but many samples. 
Since the p-value makes no distinction between these two cases, we need an \emph{effect size} to determine the size of this significant difference. 

Many effect sizes can be quite difficult to interpret, and so McGraw and Wong designed the Common Language Effect Size (CLES)~\cite{mcgraw_common_1992}. 
If we compare two populations $A$ and $B$ and have concluded that the alternative hypothesis that the median of population $A$ is greater than $B$, then the CLES describes the probability that any random observation from $A$ is greater than a sample from $B$. 
With random samples $X_A$ from $A$ and random samples $X_B$ from $B$, the CLES is defined as in Eq.~\ref{eq:a-measure}, where the 0.5 factor is a tie-breaker as described in the work by Vargha and Delaney~\cite{vargha_critique_2000}.
\begin{equation}\label{eq:a-measure}
    A(X_A, X_B) = P(X_A > X_B) + 0.5P(X_A = X_B)
\end{equation}

%% file: content/related-work/auto-tuning.tex
\section{Autotuning}\label{sec:autotuning}
In this section, we will give a high-level overview of the concepts 
of the various optimization techniques that are used in empirical auto-tuning.
We will investigate two main groups of optimizations. 
First is the model-based techniques that train a predictive surrogate model offline and then predict online. Second is the sequential model-based techniques that incorporate their model into the sample collection process.

\subsection{Random Forest \& Model-based optimization}
The problem of model-based auto-tuning and related fields is the development of an approximation function A such that a solution $a$ approximates the true solution $p$.
We use this approximation function as a \emph{surrogate} model as a substitute for running the real program. This allows us to explore the state space much more efficiently if the runtime of the surrogate model is sufficiently faster than running the real program. If the surrogate model's characteristics w.r.t to the measurement function is sufficiently similar to the original program, then the model's maximum/minimum result will ideally be very close or identical with the maximum/minimum result of the original program. We can thus use this model to reduce the total search time to find the optimal solution.

    


When designing such a surrogate model Bergstra \emph{et al.} opted for Boosted Decision Trees as models~\cite{bergstra_machine_2012}. However, \emph{bagging} where we average the output of a large ensemble of different models, is also a very effective alternative. In 2001 Breiman~\cite{breiman_random_2001} combined this bagging technique with a random selection of features to create the \emph{Random Forest} model. By varying the predictors and averaging across a 
collection of different decision trees, we 
increase the generalization performance and make the model less susceptible to noise.

\subsection{Sequential Model-Based Optimization (SMBO)}
The model-based approaches from the previous section are based on a two-stage process, with sample collection and then training and prediction. However, the samples collected for the training set greatly impact the performance of the models. We will therefore now discuss \emph{Sequential Model-Based Optimization}(SMBO). The key idea is to use our surrogate model to more intelligently collect training samples that are useful for training a precise model. 

\subsubsection{Bayesian Optimization}
The most common type of SMBO is based on \emph{Bayesian Optimization,} where we use a \emph{Gaussian Process}(GP) to find the most interesting training samples. Bergstra et al. used this method in their HyperOpt library~\cite{bergstra_hyperopt_2013, bergstra_algorithms_2011}. They also developed Tree-Parzen Estimators(TPE) as a less computationally expensive alternative. The selection criteria for the next sample in an SMBO process varies, however the most common strategy is \emph{Expected Improvement} where we collect the samples that we \emph{predict} will perform the best, given our current knowledge. There are also various other sampling strategies, including Upper Confidence Bounds and Thompson Sampling. 

    


\subsubsection{Genetic Algorithms} Among the metaheuristic optimization algorithms van Werkhoven~\cite{van_werkhoven_kernel_2019} showed that \emph{Genetic Algorithms} performs well for autotuning.

Genetic algorithms~\cite{sivanandam_genetic_2008} can also be viewed as a type of SMBO in the way that it performs crossover and mutation on previous generations to create more sophisticated solutions. Genetic algorithms start with a random population of samples and then iteratively alter this population towards better solutions. We can describe the process in 5 main steps:
\begin{enumerate}
\item Chromosomes(configurations) are randomly generated to form a population.
\item Chromosomes evaluated using measurement function.
\item The best chromosomes are kept, the rest discarded.
\item New population is generated by crossover and mutation.
\item Repeat until max iterations.
\end{enumerate}
Here, the crossover is an alteration of the configurations by combining parts of the configurations from the best chromosomes. The idea is then that a combination of good features will hopefully lead to an even better \emph{offspring}. One way to implement this is by taking half of the variables from configuration A and the other half of variables from configuration B when creating their offspring C.
Additionally, to avoid getting stuck in local extrema we add a \emph{mutation} operator that introduces stochasticity into the process. The mutation operator will with a low probability randomly change the variables.

%% file: content/related-work/related-work.tex
\section{Related work}\label{chap:related-work}
\input{content/related-work/related-fields}

%% file: content/related-work/related-fields.tex

\renewcommand{\thefootnote}{\fnsymbol{footnote}}
\begin{table*}[]
    \centering
    \caption{Overview of previous experimental designs for empirical optimizations.
    }
    \begin{minipage}{\textwidth}
    \centering
    \resizebox{\textwidth}{!}{%
    \begin{tabular}{llllll}
        \toprule
        Author & Samples/Experiments/Evaluations~\footnote{The number of evaluations of the final solution, to compensate for runtime variance.} & Significance test & Research field & Algorithms \\
        \midrule
        Hutter et al.~\cite{hutter_sequential_nodate} & 30-300 Min / 25 / 1000 & Mann-Whitney U & AlgConf & SMAC, ROAR, TB-SPO, GGA(GA) \\
        Eggensperger et al.~\cite{eggensperger_towards_nodate} & Varies\footnote
        {\label{noteB}No common value for all benchmarks or in the case of van Werkhoven, no common value for all algorithms} (50 to 200) / 10 / n/a & Unpaired t-test & AlgConf & BO TPE, SMAC, Spearmint\\
        Falkner et al.~\cite{falkner_bohb_2018} & Varies\footref{noteB} / Varies / Varies & n/a & AlgConf & RS, BO TPE, BO GP, HB, \\ &&&& HB-LCNet and BOHB \\
        Snoek et al.~\cite{snoek_practical_2012} & Varies\footref{noteB} (1-50,1-100) / 100 / n/a & n/a & HypOpt & BO GP, Grid search\\
        Bergstra et al.~\cite{bergstra_algorithms_2011} & 230 / 20 / n/a & n/a & HypOpt & RS, BO TPE, BO GP, Manual \\
        Bergstra et al.~\cite{bergstra_random_nodate} & 1-128
        / 256-2
        / n/a & n/a & HypOpt & RS, Grid Search(GS) \\
        
        Bergstra et al.~\cite{bergstra_machine_2012} & 10-200
        / n/a / n/a & n/a & HypOpt & Boosted Regression Trees, \\ &&&& GS, Hill Climbing\\  
        
        Falch and Elster~\cite{falch_machine_2017} & 100-6000
        / 20 / n/a & n/a & Autotuning & NN, SVR, Regression Tree \\
        
        van Werkhoven~\cite{van_werkhoven_kernel_2019} & Varies\footref{noteB}
        / 32 / 7 & n/a & Autotuning & Many Metaheuristic Methods\\
        
        Willemsen et al.\cite{willemsen_bayesian_2021} & 20-220 / 35 / n/a & n/a & Autotuning & BO, RS, SA, MLS and GA \\
        
        Ansel et al.~\cite{ansel_opentuner_2014} & Varies\footref{noteB} / 30 / n/a & n/a & Autotuning & Multi-armed bandit, Manual \\
        
        Nugteren et al.~\cite{nugteren_cltune:_2015} & Varies\footref{noteB} (107 or 117)/ 128 / n/a & n/a & Autotuning & RS, SA, PSO \\
        
        Akiba et al.~\cite{akiba_optuna_2019} & Varies\footnote
        {The samples sizes vary, due to pruning techniques.} / 30 / n/a & "Paired MWU"
        & Autotuning &  RS, HyperOpt, SMAC3, \\ &&&& GPyOpt, TPE+CMA-ES\\
        
        Grebhahn et al.~\cite{grebhahn_predicting_2019} & 50, 125 / Unclear~\footnote{The paper mentions 15000 experiments, however it does not state how many experiments per algorithm/benchmark.} / n/a & "Wilcox test"
        & SBSE & RF, SVR, kNN, CART, KRR, MR \\
        
        \midrule
        Tørring & 25-400
        / 800-50
        / 10 & Mann-Whitney U & Autotuning & RS, BO TPE, BO GP, RF, GA \\
        \bottomrule
    \end{tabular}
    }
    \end{minipage}
    \label{tab:related-work-overview-table}
\end{table*}

\renewcommand*{\thefootnote}{\arabic{footnote}}

\input{content/related-work/background-related-fields}
\input{content/related-work/overview-related-fields}

%% file: content/related-work/background-related-fields.tex
In this section, we will give a high-level overview of the related research fields and then investigate each of these related research fields for related works.

\subsection{Hyperparameter optimization in ML}
Since autotuning is simply an application of mathematical optimization for program optimization, we can see strong similarities with \emph{hyperparameter optimization}. 
The machine-learning(ML) community often creates models that contain \emph{hyperparameters} that greatly impact the performance of the models. 
The community has therefore performed extensive research in how to most efficiently optimize these hyperparameters. 
Hyperparameter optimization therefore focuses on the \emph{optimization of models,} where the measurement function $M$ is usually the loss $L$ of the model.

In the field of hyperparameter optimization, J. Bergstra has provided several seminal works. 
He presented results that showcased Random Search as superior to Grid Search for hyperparameter optimization~\cite{bergstra_random_nodate} and that 
both BO GP and BO TPE algorithms outperform Random Search~\cite{bergstra_algorithms_2011}. 
In 2012 Bergstra also published an autotuning technique based on machine learning with boosted regression trees~\cite{bergstra_machine_2012}, before publishing a hyperparameter optimization library, HyperOpt, based on these techniques~\cite{bergstra_hyperopt_2013}.
Snoek et al. also presented their application of Bayesian Optimization with Gaussian Processes in 2012~\cite{snoek_practical_2012}.

\subsection{Algorithmic Configuration}
A generalization of the hyperparameter optimization field includes \emph{Algorithmic Configuration,} where the field focuses on general optimization of algorithms. 
In this field, the measurement function $M$ could be any meaningful function for the performance of the algorithm. 
Frank Hutter's AutoML group has contributed significant research in this community~\cite{hutter_sequential_nodate, eggensperger_towards_nodate, falkner_bohb_2018}. This includes developing the BOHB algorithm that combines Hyperband with TPE-based Bayesian Optimization~\cite{falkner_bohb_2018}.





\subsection{Comparative studies}
The most relevant comparative study is the work by Falch and Elster where they published an Autotuner, AUMA, based on a feedforward neural network in 2017~\cite{falch_machine_2017}. 
The autotuner was designed specifically to autotune ImageCL and OpenCL kernels. 
The proposed technique showed promising results.

In 2019, van Verkhoven published comprehensive results comparing various metaheuristic optimization techniques, and these results indicate that Genetic Algorithms performs well among the metaheuristic optimization techniques~\cite{van_werkhoven_kernel_2019}. 
These results also focus on OpenCL kernels. 
The results are based on 32 experimental runs and a single sample size for each algorithm.
The results are therefore a good indication for future research, yet more research is needed to compare against other state-of-the-art techniques. The work has since been extended by Willemsen et al.\cite{willemsen_bayesian_2021}. Their 2021 work compares Bayesian Optimization to previous search techniques implemented in Kernel Tuner. The results indicate that Bayesian Optimization outperforms the other search techniques for most kernels and sample sizes. 

\subsection{Frameworks}
Rasch et al. published ATF: A generic autotuning framework in 2019~\cite{rasch_atf_2019}. 
The framework was compared against the older OpenTuner framework~\cite{ansel_opentuner_2014} and CLTune framework by C. Nugteren and V. Codreanu~\cite{nugteren_cltune:_2015}. 
CLTune~\cite{nugteren_cltune:_2015} compared several of the state-of-the-art techniques at the time, with an experiment size of 128. 
They performed two experiments on two benchmarks, with sample sizes of 107 and 117, respectively. 
The results give a strong indication that Simulated Annealing(SA) and Particle Swarm Optimization (PSO) outperforms Random Search, however the superiority of PSO and SA depends on the benchmark at hand. 
The experimental size of this study could be sufficient to make such conclusions significant, however the authors do not provide any significance test. 

Akiba et al. provides a very thorough experimental design and analysis with significance tests when comparing their hyperparameter optimization framework Optuna~\cite{akiba_optuna_2019} against HyperOpt~\cite{bergstra_hyperopt_2013}, SMAC3~\cite{hutter_sequential_nodate} and other frameworks. 
However, the study compares the performance of the frameworks and not the underlying algorithms in the framework. 
The study is thus a very good indicator of the performance of the Optuna framework yet does not provide any significant general conclusions about the algorithms. 

\subsection{Search-based Software Engineering}
The software engineering community has also invested significant research into finding techniques for optimal application-level configurations. 
Among the most recent work is an excellent empirical study by Grebhahn et al.~\cite{grebhahn_predicting_2019}.
The authors evaluated several popular techniques in Software Engineering, including k-Nearest Neighbours, Support Vector Regression and Random Forests for creating surrogate performance models. 
These models can then be used to predict optimal application-level configurations. 


%% file: content/related-work/overview-related-fields.tex
\subsection{Overview of related studies}

Given all these related works, we have compiled a Table of the experimental details and algorithms for each of the most influential works that we have found. In Table~\ref{tab:related-work-overview-table} we present the experimental details of the Related works we found in our survey. 
The rows in the table also include the experimental details for our design as we will discuss in Section~\ref{sec:experimental-design}. Some tools presented as algorithms here are \emph{frameworks} and thus employ more techniques than the core search algorithms they are based on. 

From analyzing this table, we can note that while the HypOpt/AlgConf field has focused on various implementations of Bayesian Optimization, the Autotuning field has focused on Metaheuristics optimization and traditional techniques, e.g. SA, PSO, GA, SVR, kNN, etc. 
We can therefore not find any modern study that compares the state-of-the-art algorithms in autotuning with the best HypOpt or AlgConf algorithms.
Very few evaluate a consistent range of sample sizes, with Bergstra's publications being the most common exception. 
Many publications also perform less than 50 experiments and do not provide any significance tests to assess the significance of their results. 
Few of the previous works provide any effect sizes for the results, either. However, some previous research has evaluated a larger range of different techniques, especially the thorough work by van Werkhoven~\cite{van_werkhoven_kernel_2019}.

%% file: content/method/method.tex

\input{content/method/setup}

\input{content/method/framework}

%% file: content/method/setup.tex
\section{Experimental Design}\label{sec:experimental-design}


For our experiments, we assume that the time it takes to run predictions using the models is insignificant compared with the cost of running the samples. This leads us to a design where we want to investigate how \emph{sample-efficient} each algorithm is. I.e. we want to compare the algorithms for how well the best predicted configuration performs, given a fixed number of samples for all algorithms.

The justification for such a design is based on case studies where large applications are autotuned~\cite{sid-lakhdar_multitask_2019}. In cases like these, the surrogate model runtimes are negligible compared to running the programs.

It also allows us to evaluate how the \emph{algorithms} perform instead of their implementation. The measured runtime of these algorithms are usually heavily dependent on implementation details, such as the programming language. 

\subsection{Distribution of samples and significance tests}\label{subsec:distribution-all}

When we performed our initial random sample experiments, we got a population of samples that was obviously non-gaussian. Neither could the populations be modeled accurately with any of the distributions in the SciPy statistics package.
Therefore, we cannot make any assumptions about the underlying distribution, so we need a non-parametric significance test. An alternative could be Bootstrapping, however that would drastically increase the time it takes to get sufficient results.

We propose to use the Mann-Whitney U(MWU) test or the Wilcoxon rank-sum test. The choice of this test is motivated by the wide-spread support of the Wilcoxon rank-sum test for these types of studies~\cite{arcuri_practical_2011, grebhahn_predicting_2019}.
For our study, we chose the significance threshold $\alpha = 0.01$. 

\subsection{Experiment size}
When choosing how many experiments to perform for each algorithm, we need to evaluate the variance between the results and how many experiments we need to achieve a sufficient p-value.


However, as we ran the algorithms for different sample sizes, we noticed that the variance in our results decreased as a function of sample size. 
We therefore seek to scale the number of experiments for each algorithm as a function of the sample sizes we are evaluating. This will allow us to run more experiments for lower sample sizes that have a higher variance, and run fewer experiments for higher sample sizes. 

With the assumption that we wanted \emph{at least} $50$ experiments for our $sample\_size = 400$ case, we performed $800$ experiments for our $sample\_size = 25$ case and scaled the number of experiments for the rest of the sample sizes similarly.




\subsection{Hyperparameters and Search Space}


We have limited our study to \emph{best guess} hyperparameters, assuming that the inherent difference between the algorithms amortizes the difference between our best guess hyperparameters and the ideal hyperparameters.

Our study includes 6 tuning parameters. The search space of the benchmarks is defined 
by the three thread dimensions with range $\{X,Y,Z\}_t = [1..16] $ and the three work group size parameters with range $\{X,Y,Z\}_w = [1..8]$. This gives a total 
of $|S| = 2\ 097\ 152$ configurations.
We initially generated the training samples for the model-based approaches using a constraint specification. 
We knew from prior knowledge that the product of our work group size parameters must not exceed 256.

Using this constraint, we only generated executable configurations for our non-SMBO methods. 
The SMBO-methods did not have any option for specifying constraint specification in their searches. 
We therefore consider the use of constraint specification a design point in which non-SMBO methods are favored. As we will show, the SMBO methods still greatly outperformed non-SMBO methods. 

\subsection{Benchmarks, Hardware and Runtime Environment}
The benchmarks for this study consist of three benchmarks from the ImageCL benchmark suite~\cite{falch_machine_2017}.
The Add benchmark consists of a simple vector addition with two vectors of size $X$.
The Harris benchmark is slightly more complex than the Add benchmark, as it involves executing the \emph{harris corner detection} algorithm. This algorithm is used to detect where corners are in images. The algorithm is performed on an image of size $X$ by $Y$ and is also easily parallelizable, like the Add benchmark.
The final benchmark is the construction of an image of size $X$ by $Y$ with intensity values according to the Mandelbrot set. This will create the classic visualization of the Mandelbrot set. This is also an easily parallelizable benchmark that lends itself to a GPU.

We ran all the benchmarks with, $X=8192, Y=8192$ which is the default value for previous internal studies on autotuning of ImageCL. 
All of our benchmarks were run on three different GPU architectures to compare the results. 
We specifically chose the RTX Titan from 2019, Titan V from 2017 and GTX 980 from Fall 2014 to compare between older and more modern architectures. 



%% file: content/method/framework.tex
\section{Autotuning framework}\label{sec:framework}
\begin{figure*}[]
    \centering
    \includegraphics[width=12cm]{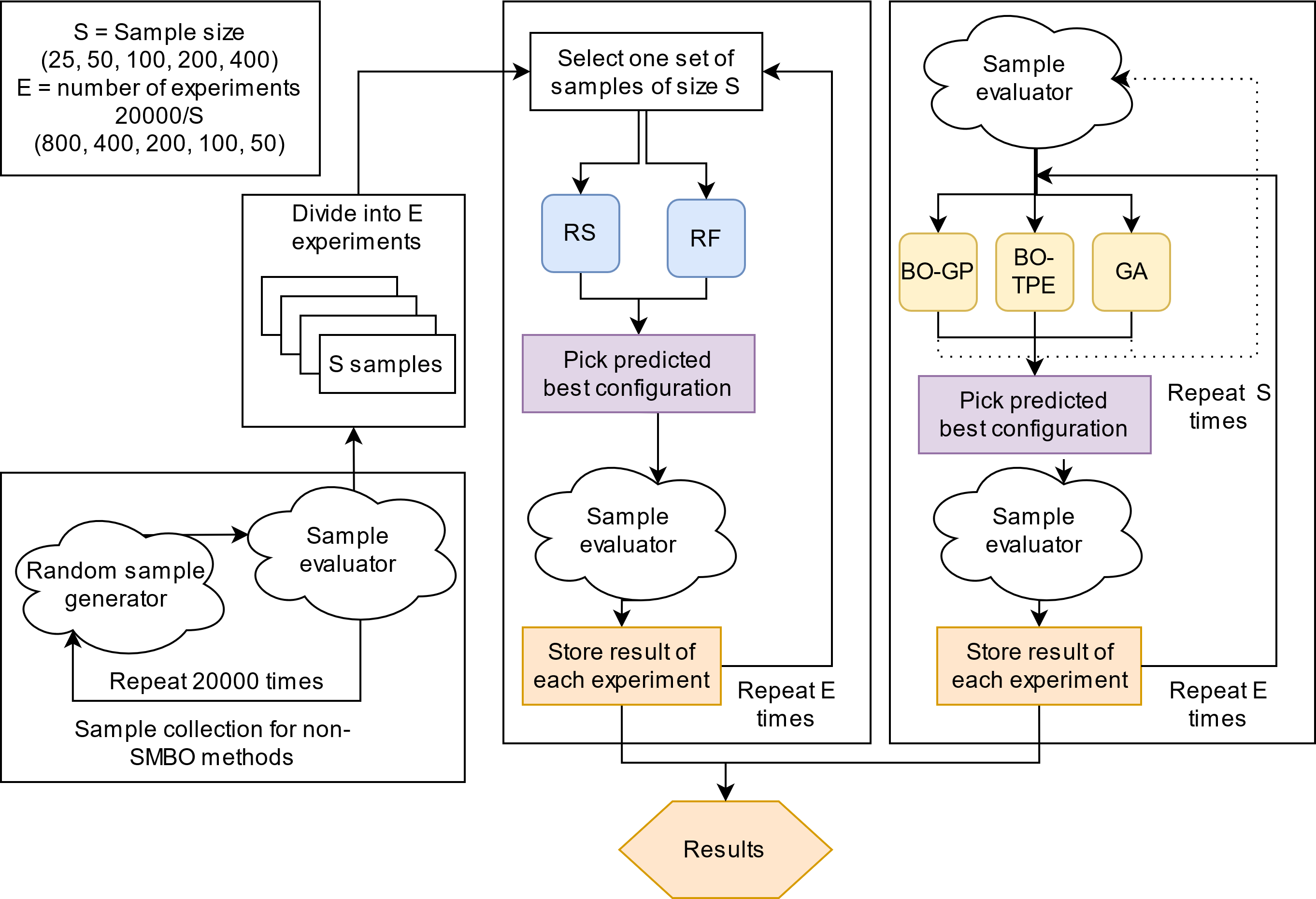}
    \caption{Experiment pipeline, including sample collection process and prediction stage.}
    \label{fig:experiment-pipeline}
\end{figure*}

In this section we will describe the implementation of our autotuning framework and techniques as illustrated in Fig.~\ref{fig:experiment-pipeline}.

\subsection{Measurement}
The total runtime of performing a computation on a GPU is heavily dependent on data transfer between the CPU's main memory and the GPUs memory via the PCIe-bus. Therefore, it is imperative that we first transfer the data and \emph{then} start the measurement timer when executing the kernel. Respectively, the timer needs to stop after the kernel has finished, but \emph{before} the data is transferred back to the host device. This ensures that we measure the actual execution time of the computation and not the additional time to transfer data between devices.

When the autotuning algorithm has terminated, we test the final sample 10 times to compensate for runtime variance. We only run the sample once during the training and sampling process to better represent real use cases and test the models for how well they handle noise in the samples.

\subsection{Implementation of the algorithms}
When implementing autotuning algorithms, the runtime of each algorithm is heavily dependent on implementation details, such as the choice of language. As discussed in Sec.~\ref{sec:experimental-design}, we focus on the sample-efficiency of the algorithms instead of runtime. 


We have implemented the following methods in our experimental frameworks:
Random Search (RS),
Random Forest Regression (RF),
Genetic Algorithms (GA),
Bayesian Optimization with Gaussian Processes (BO-GP),
Bayesian Optimization with Tree-Parzen Estimators(BO-TE).

For our non-SMBO approaches, we streamline the experimental sample collection process by creating a dataset of 20 000 samples in one go for each architecture and benchmark. We can then subdivide the samples for each sample size and experiment.

For the case of
Random Search (RS), 
we simply select the minimum runtime from the collection of $S$ samples for the given experiment. We then collect the minimum for all $E$ experiments in the sample size. 

For model-based approaches like Random Forest (RF), we train the models with the subset of size $S - 10$ for each experiment and then run the top 10 predictions. The top performing prediction is then stored as the output. We repeat this process for all $E$ experiments. 
The RF model is implemented using the \emph{sk-learn} RandomForestRegressor model. 
Configurations are run 10 times to compensate for runtime variance and to decrease noise in our experimental results.

Bayesian Optimization with Gaussian Processes is implemented using the Scikit-optimize's gp\_minimize function. The acquisition
functions is defined as the Expected Improvement.
Initialization uses 8\% of the samples, and the remaining 92\% are used as prediction samples in the search.

For the TPE variant of BO we used the Hyperopt library by Bergstra et al.~\cite{bergstra_hyperopt_2013}. The implementation is very similar to the GP-variant, where we specify a search range and the number of evaluations, as well as the objective/measurement function. The only limitation of this library compared to gp\_minimize is the inability to specify the balance of random samples to model-driven samples.

To make our study as comparable as possible we based our Genetic Algorithm implementation on the implementation that van Werkhoven used in their study~\cite{van_werkhoven_kernel_2019}. We have thus only made minor changes to make the implementation compatible with our experimental framework. 

%% file: content/results/results.tex
\section{Results and Discussion}\label{chap:results}
\begin{figure}
    \centering
    \includegraphics[width=1.05\linewidth]{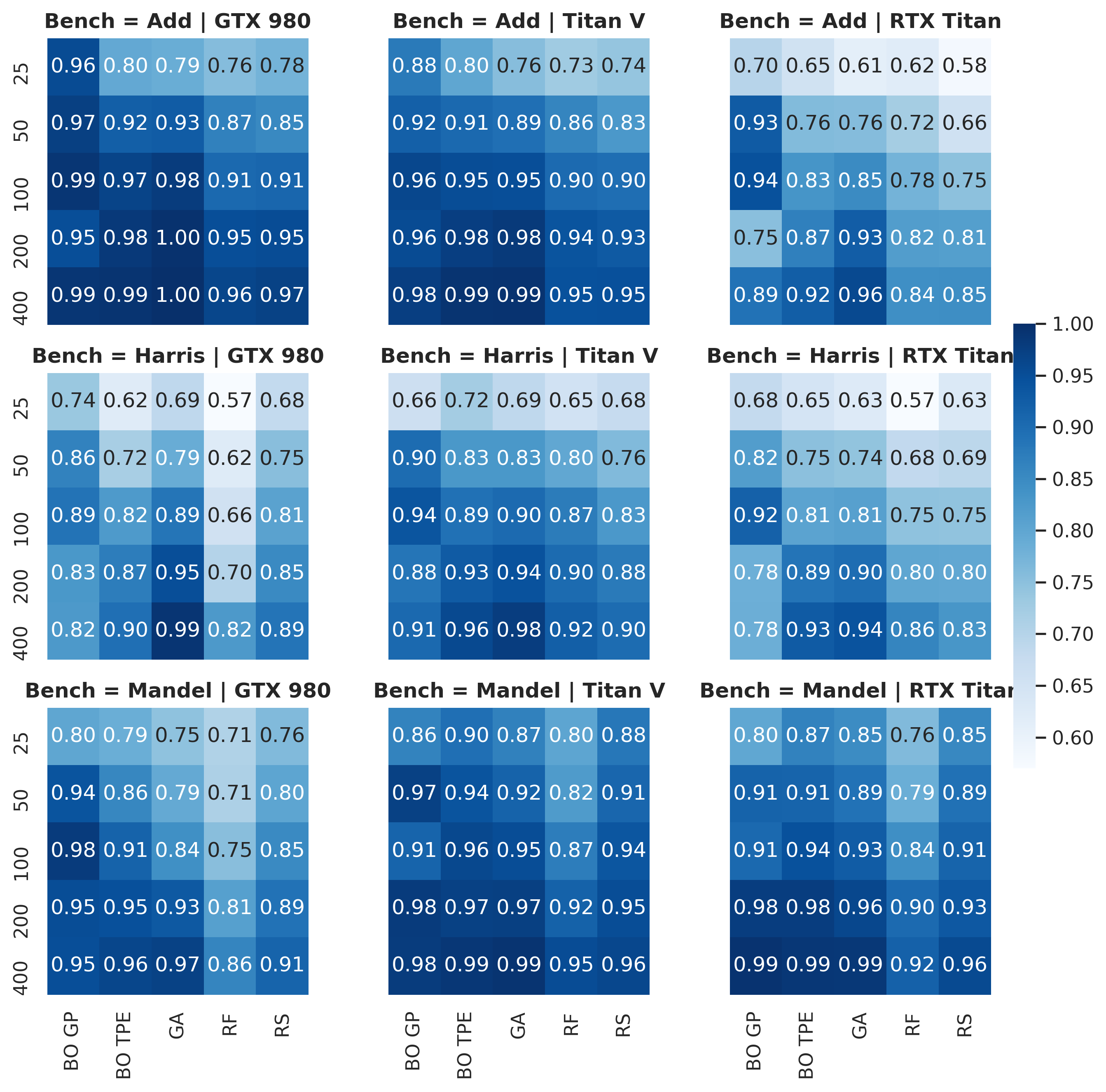}  
    \caption{Percentage of optimum performance for each algorithm on each benchmark and architecture.}
    \label{fig:facet_grid_optimum}
\end{figure}

\begin{figure}[ht]
    \centering
    \includegraphics[width=\linewidth]{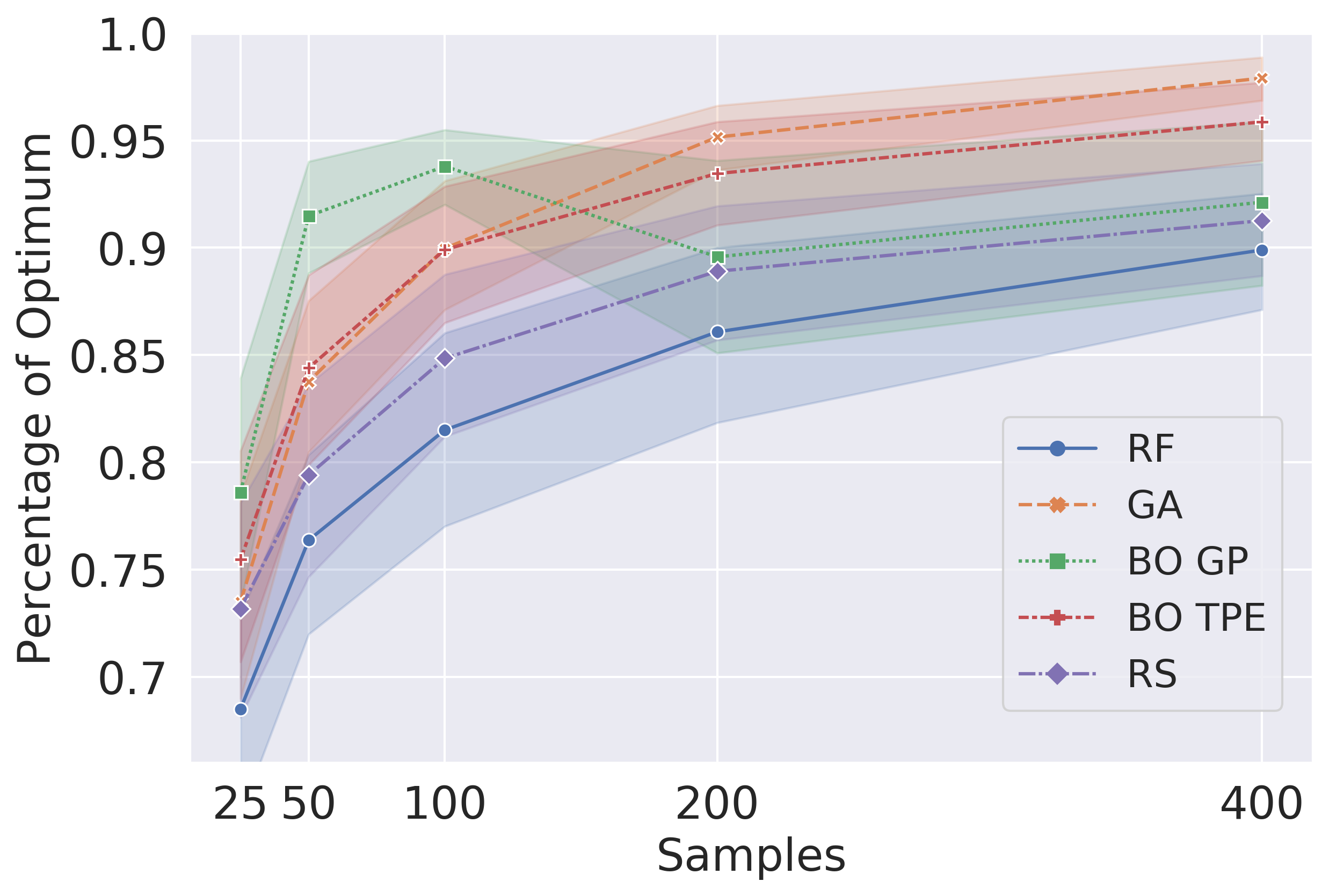}
    \caption{Mean and confidence interval of percentage of optimum performance across all benchmarks and architectures.}
    \label{fig:optimum_plot}
\end{figure}

\begin{figure*}[ht]
\begin{subfigure}{.5\textwidth}
  \centering
  \includegraphics[width=\linewidth]{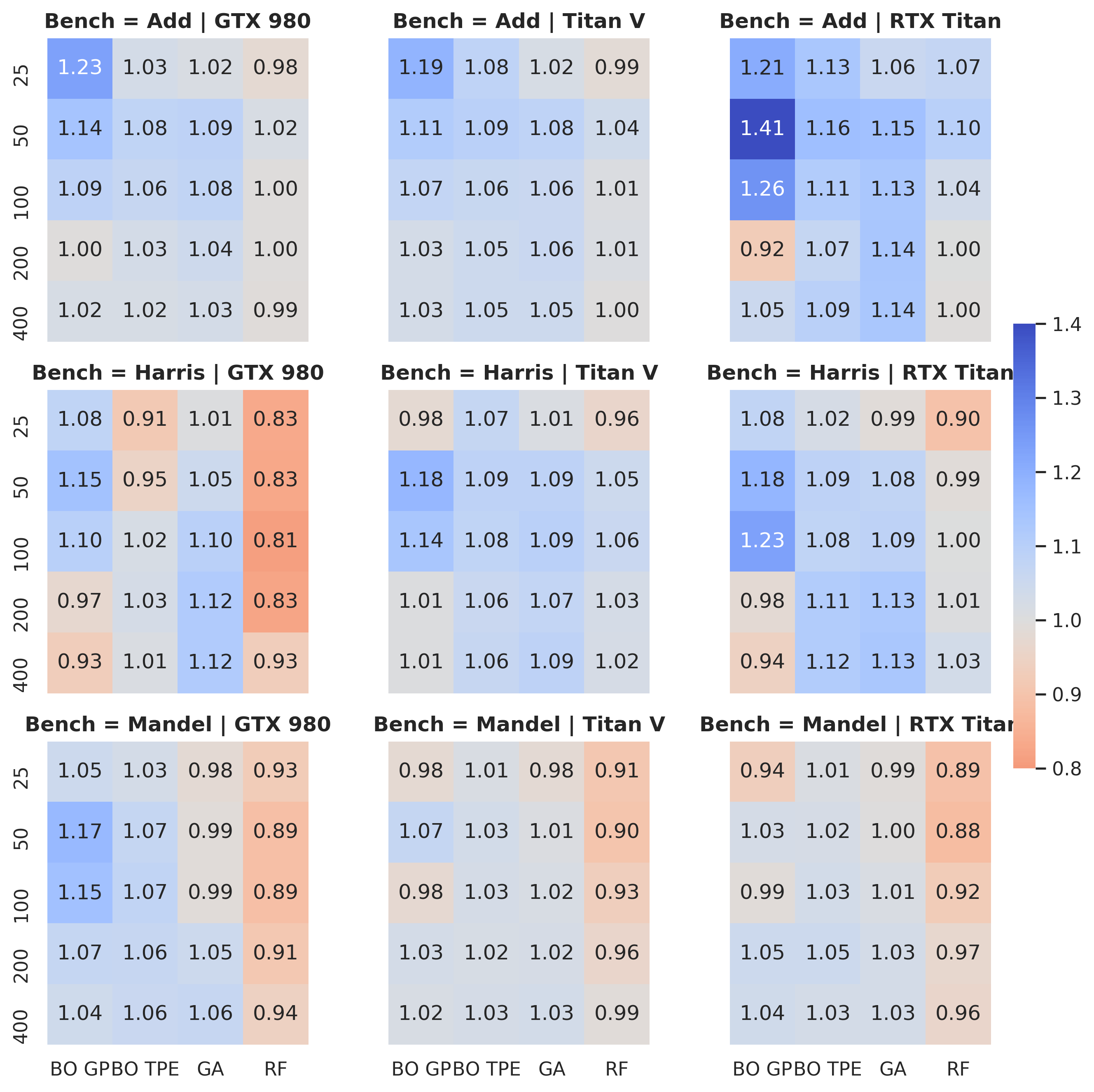}  
  \caption{Median speedup over Random Search}
  \label{fig:facet_grid_median}
\end{subfigure}
\begin{subfigure}{.5\textwidth}
  \centering
  \includegraphics[width=\linewidth]{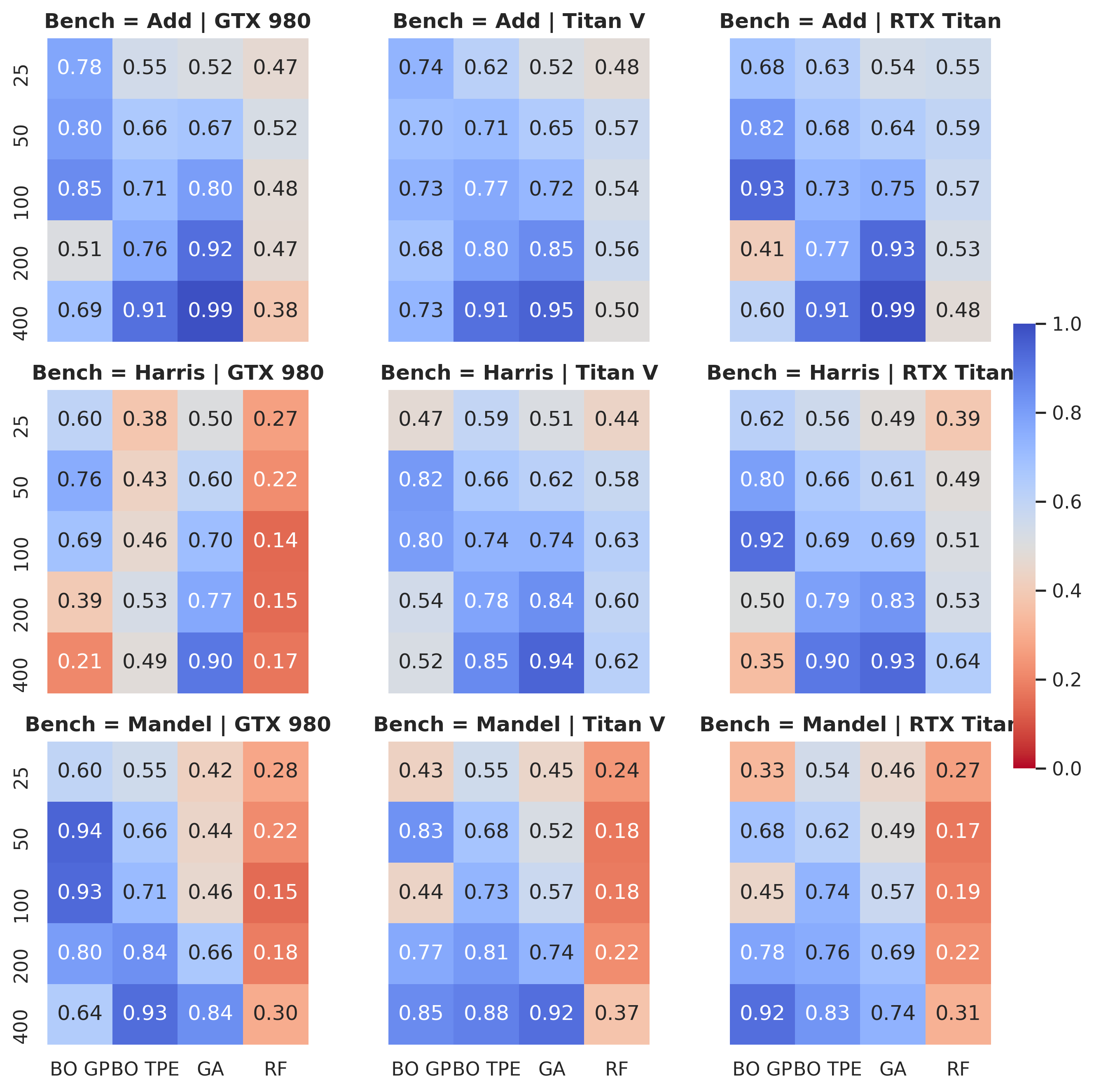}
  \caption{Common Language Effect Size over Random Search}
  \label{fig:facet_grid_cles}
\end{subfigure}
\caption{Median speedup over Random Search and CLES over Random Search 
}
\label{fig:facet_grid}
\end{figure*}

This section will present the results from all of our roughly 3 019 500 samples\footnote{3 SMBO algorithms, [25, 50, 100, 200, 400] samples per algorithm, [800, 400, 200, 100, 50] experiments + RS/RF Samples and RF predictions for 3 benchmarks on 3 architectures.} across various algorithms, benchmarks and architectures. In addition, we will present the general trends and data across our results in this section. 

One would typically want to choose the autotuning algorithm, which is most likely to give us the best result. This can be done by comparing the median of all our runs of the various algorithms under question.

When providing the autotuning algorithms with more samples, we often want to see how close they get to the optimum solution. Fig.~\ref{fig:facet_grid_optimum} therefore provides how close the median solution of each algorithm's run on a benchmark and architecture is to the study's optimum solution. We visualize the results from these heatmaps as an aggregate line plot of the mean value of the medians in Fig.~\ref{fig:optimum_plot} with the confidence interval generated from the heatmap values.
To better understand how all the algorithms behave compared to Random Search, we also provide a range of heatmaps that show the relative speedup of each algorithm compared with Random Search, Fig.~\ref{fig:facet_grid_median}. 
Fig.~\ref{fig:facet_grid_cles} is the corresponding CLES 
plot of the values 
, i.e. the probability of the algorithm's solution outperforming Random Search. 

We view all cases statistically significant ($\alpha=0.01$) where a given algorithm's median performance differs by more than 1\% compared with other algorithms.
This is statistically significant under the MWU test, since our number of experiments are sufficiently high (for sample size 25 to 400, with 800 to 50 experiments, depending on the sample size).


\subsection{Percentage of Optimum solution}
The results in Fig.~\ref{fig:facet_grid_optimum} and \ref{fig:optimum_plot}, show that all the benchmarks and architectures share a few common trends. 
We can see that BO GP outperforms most other algorithms for low and mid-range sample sizes from 25 to 100 for most of the benchmarks and architectures. From sample size 100 to 200 BO GP has a decline in performance, potentially due to overfitting~\cite{makarova_overfitting_2021}. This effect varies between benchmarks, yet is a consistent trend for BO GP, while all other algorithms have strictly increasing performance as a function of sample size. The extent of this effect is likely influenced by the dimensionality and cardinality of the search space, in addition to the sample sizes we are investigating. The benchmark also seems to affect this behavior, with BO GP providing peak performance on the Harris benchmark at sample size 100. In contrast, for the Add benchmark, the peak varies between architectures. BO GP, on average, provides better performance for the highest sample size for the Mandelbrot benchmark.

For sample sizes of 200 and 400, GA outperforms all other algorithms for most benchmarks and architectures. 
BO TPE also performs well for many of these benchmarks and architectures in the same sample size range. BO TPE performs well for most sample sizes, yet is often outperformed by BO GP or GA in their respective sample size ranges. 
The Non-SMBO RF method often performs worse than RS and never outperforms all the other methods. 


\subsection{Speedup over Random Search}
When we compare the results of each algorithm against Random Search in Fig.~\ref{fig:facet_grid_median}
, we can see a few general trends. 
Compared with RS, the potential relative gain of using advanced search techniques like BO GP and BO TPE is greatest for smaller sample sizes.
Using BO GP or BO TPE for sample sizes from 25 to 100 generally gives us 10-40\% better performance than simply using RS. However, some combination of benchmarks and architectures give less speedup, e.g. Mandelbrot on Titan V and RTX Titan.


For higher sample sizes like 200 and 400, GA generally outperforms the other algorithms, with BO TPE also performing well. However, the potential speedup over RS is less in this range, from 3-14\%, depending on the benchmark and architecture. We would expect this trend to continue as the sample size increases and a larger portion of the search space is evaluated by RS.



\subsection{Probability of outperforming Random Search (CLES)}
In Fig.~\ref{fig:facet_grid_cles}
we present a plot of each algorithm's Common Language Effect Size over Random Search for all benchmarks and architectures. 
These figures can essentially be interpreted as \emph{probability} plots of how often an algorithm will outperform the alternative. 
Since our alternative algorithm in these plots is RS, these plots show the percentage of experiments for a given algorithm that outperformed random search. 

We can see similar trends in these plots 
where BO GP performs well for low to mid-range sample sizes and GA performs well for high sample sizes. BO TPE also appears to be a good balance across the entire sample range, often outperforming GA for low sample sizes and BO GP for high sample sizes. 
While the potential for performance improvements over RS is greatest for lower sample sizes, we can observe that the algorithms more consistently outperform RS for higher sample sizes.


%% file: content/conclusion.tex
\section{Conclusion and Future work} \label{chap:conclusion}


Optimizing codes for modern computing systems with multi-core CPUs and GPUs that update their architecture often, is becoming increasingly challenging. Autotuning addresses this by optimizing parameterized code to the targeted hardware by \textit{searching} for the optimal set of parameters, including addressing memory hierarchies, etc. In this paper, we analyzed through experimental results some of the most popular search techniques used in autotuning and how they compare for various benchmarks, architectures and sample sizes. 



Our study compared the search algorithms Random Search (RS), Random Forest Regression (RF), Genetic Algorithms (GA), Bayesian Optimization with Gaussian Processes (BO GP) and Bayesian Optimization with Tree-Parzen Estimators (BO TPE).
The algorithms were compared using three image-based GPU kernels on three GPU architectures, including the RTX Titan, Titan V and GTX 980. 
While the largest differences in autotuning results were between sample sizes, the benchmark and architecture were also shown to significantly impact the results.

Our results showed that advanced search techniques outperform random search to a larger extent for low sample sizes between 25 and 100 samples. 
For higher sample sizes from 200 to 400 samples, advanced search techniques more consistently outperform random search. Note that no single search technique, among those tested, outperforms all other techniques for all sample sizes. BO GP and BO TPE perform well for low sample sizes from 25 to 100, and GA performs best for higher samples sizes from 200 to 400.

\input{content/future-work}

%% file: content/future-work.tex
\subsection{Current and Future work}
Our results provide great motivation for future work into better understanding how the relative performance of how search algorithms change as functions of the sample size, benchmarks and architectures.
Current work thus includes testing a wider range of benchmarks~\cite{sund_BAT_2021, bjertnes_ls-cat_2021}, architectures and search algorithms for a wider range of sample sizes. This will naturally require even more computational resources than we already used for our study. Comparing our selection of algorithms 
against HyperBand(HB) and Bayesian Optimization HyperBand(BOHB)~\cite{falkner_bohb_2018} as well as Deep-learning based methods~\cite{bjertnes_autotuning_2021} is of special interest. Investigating the effect of different input data sets to the benchmarks could also provide insightful results. 

%% file: content/acknowledgements.tex
\section*{Acknowledgements}
The authors would like to acknowledge 
NTNU for the PhD stipend and support of our HPC-Lab that facilitated the development of this project. 
The authors would also like to acknowledge 
Zawadi Berg Svela, 
Tor Andre Haugdahl 
and Jan Christian Meyer 
for their valuable feedback. The second author would also like to thank the Center for Geophysical Forecasting (SFI CGF) at NTNU and RCN (NFR proj. no. 309960) for their support during this project.


%% file: content/generated/references.tex
\bibliographystyle{unsrt}
\bibliography{mylib, references}